\documentclass[12pt]{article}
\usepackage{epsf}
\usepackage{graphicx}

\begin{document}

\title{Secure exchange of information by synchronization of neural  
networks}  
\author{Ido Kanter$^1$, Wolfgang Kinzel$^2$ and Eran Kanter$^1$}    
\date{}    
\maketitle    
\begin{center}    
     
(1) Department of Physics, Bar Ilan University, 52900 Ramat    
Gan, Israel

(2) Institut f\"{u}r Theoretische Physik,    
Universit\"{a}t W\"{u}rzburg, Am Hubland, D-97074 W\"{u}rzburg,    
Germany    
     
\date{3. 9.  2001}    
\end{center}            
    
\begin{abstract}    
  A connection between the theory of neural networks and    
    cryptography is presented. A new phenomenon, namely synchronization    
    of neural networks is leading to a new method of exchange of    
    secret messages. Numerical simulations show that two artificial    
  networks being trained by Hebbian learning rule on their mutual    
  outputs develop an antiparallel state of their synaptic weights. The    
  synchronized weights are used to construct an ephemeral key    
    exchange protocol for a secure transmission of secret data. It is    
  shown that an opponent who knows the protocol and all details of any    
  transmission of the data has no chance to decrypt the secret    
  message, since tracking the weights is a hard problem compared to    
  synchronization. The complexity of the generation of the secure    
  channel is linear with the size of the network.    
\end{abstract} 
   
\centerline{PACS numbers: 87.18.Sn,89.70.+c}     
\vspace{0.5cm}

The ability to build a secure channel is one of the most challenging    
fields of research in modern communication.  Since the secure channel    
has many applications, in particular for mobile phone, satellite and    
internet-based communications, there is a need for fast, effective and    
secure transmission protocols \cite{1}. Here we present a novel    
principle of a cryptosystem based on a new phenomenon which we    
observe for artificial neural networks.    
    
The goal of cryptography is to enable two partners to communicate    
over an insecure channel in such a way that an opponent cannot    
understand and decrypt the transmitted message.  In a general    
scenario, the message is encrypted by the sender through a key    
$E_k$ and the result, the ciphertext, is sent over the channel. A    
third party, eavesdropping on the channel, should not determine    
what the message was.  However, the recipient who knows the    
encryption key can decrypt the ciphertext using his private key    
$D_k$.     
    
In a {\it private--key} system the recipient has to    
agree with the sender on a secret key $E_k$, which requires a    
hidden communication prior to the transmission of any message.  In    
a {\it public--key} system, on the other side, the key $E_k$ is    
published and a hidden communication is not necessary.    
Nevertheless, an opponent cannot decrypt the transmitted message    
since it is computationally infeasible to invert the encryption    
function without knowing the key $D_k$. In a {\it    
key-exchange protocol}, both partner start with private keys and    
transmit -- using a public protocol -- their encrypted private    
keys which, after some transformations, leads to a common secret    
key.  In most applications a public-key system is used which is    
based on number theory where the keys are long integers \cite{1,2}.    
    
In this report we suggest a novel cryptosystem. It is a    
key-exchange protocol which does neither use number theory nor a    
public key, but it is based on a learning process of neural networks:     
The two participants start from a secret set of vectors    
$E_k(0)$ and $D_k(0)$ without knowing the key of their partner. By    
exchanging public information the two keys develop to a common time    
dependent key $E_k(t)=-D_k(t)$, which is used to encrypt and decrypt a    
given message. An opponent who knows the algorithm and observes any    
exchange of information is not able to find the keys $E_k(t)$ and    
$D_k(t)$.  Our method is based on a new phenomenon presented here:    
Synchronization of neural networks by mutual learning \cite{3a}.    
    
Simple models of neural networks describe a wide variety of phenomena    
in neurobiology and information theory \cite{3,4,5}. Artificial neural    
networks are systems of elements interacting by adaptive couplings    
which are trained from a set of examples. After training they function    
as content addressable associative memory, as classifiers or as    
prediction algorithms.     
    
In this report we present a new phenomenon: Two feedforward networks    
can synchronize their synaptic weights by exchanging and learning    
their mutual outputs for given common inputs.  Surprisingly,    
  synchronization is fast; the number of bits required to achieve    
  perfect alignment of the weights is lower than the number of    
  components of the weights.  After synchronization, the synaptic    
weights define the common time dependent private key $E_k(t)=-D_k(t)$.    
With respect to possible applications we find that first, tracking the    
weights of one of the networks by the opponent is a hard problem.    
Although we were not able to find a mathematical proof, our simulations,    
in addition to arguments based on analytic results    
on neural networks,     
give clear evidence that our key exchange protocol is secure \cite{ref2}.     
Second, the complexity of our cryptosystems scales linearly with the    
size of the network (=number of bits of the keys). In summary,    
from this new biological mechanism  one can construct      
efficient encryption systems using keys which change permanently.      
    
This phenomenon, as well as the corresponding applications in    
crypto\-graphy, can be extended to a system of several partners    
communicating with each other, as well as to other tasks relying on    
a secure channel \cite{ref2}.  Since synchronization is a subject of recent    
research in neuroscience    
too \cite{6,7,8}, we believe that our    
bridge between the theory of neural networks and cryptography may    
help to understand communication between parts of biological neuronal    
or genetic networks.

\begin{figure}[ht]     
\centering \includegraphics[width=8cm]{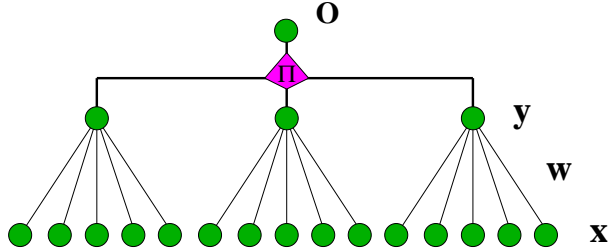}    
\caption{Architecture of the networks: $3 N$ input units $x$ are      
transmitted by three weight vectors $w$ to three hidden units $y$.      
The final output bit $O$ is the product of the hidden units.      
         \label{par}}     
\end{figure}

In the following we introduce and investigate a simple model which    
shows the properties sketched above. The architecture used by the    
recipient and the sender is a two-layered perceptron, exemplified here    
by a parity machine (PM) with $K$ hidden units. More precisely, the    
size of the input is $KN$ and its components are denoted by    
$x_{kj},~k=1,~2,~...,~K$ and $j=1,~...,~N$. For simplicity, each input    
unit takes binary values, $x_{kj} =\pm1$.  The $K$ binary hidden units    
are denoted by $y_1,~y_2,~...,~y_K$. Our architecture is characterized    
by non-overlapping receptive fields (a tree), where the weight from    
the j$th$ input unit to the k$th$ hidden unit is denoted by $w_{kj}$,    
and the output bit $O$ is the product of the state of the hidden units    
(see Fig. \ref{par}). For simplicity we discuss PMs with three hidden    
units $K=3$.  We use integer weights bounded by $L$, i.e.  $ w_{kj}$    
can take the values $-L,~-L+1,~...,~L$.    
    
The secrete information of each of the two partners are the initial    
values for the weights, $w_{kj}^S$ and $w_{kj}^R$, for the sender and    
the recipient, respectively.  It consists of $6N$ integer numbers, $3N$    
of the recipient and $3N$ of the sender. Sender and recipient do not    
know the initial numbers of their partners, which are used to    
construct the common secret key.    
     
Now each network is trained with the output of its partner.    
At each training step, for the synchronization as well as for the     
 encryption/decryption step, a new common public  input vector $(x_{kj})$     
is needed for both the sender and the recipient.     
For a given input, the output is calculated in the following two steps.     
In the first one, the state of the three hidden units, $y^{S/R}_k,      
k=1,2,3$, of the sender and the recipient are determined     
from the corresponding fields     
\begin{equation}     
y^{S/R}_k = \mbox{sign} \lbrack      
\sum\limits^N_{j=1} \; w^{S/R}_{kj} \; x_{kj} \rbrack     
\end{equation}     
    
\noindent In the case of zero field,      
$\sum w^{S/R}_{kj} \; x_{kj} =0$, the sender/recipient sets      
the hidden unit to $1/-1$.    
In the next step the output $O^{S/R}$ is determined by the      
product of the hidden units,     
$O^{S/R}= y^{S/R}_1  \, y^{S/R}_2  \, y^{S/R}_3$.

The sender is sending its output (one bit) to the recipient, the    
recipient is sending its output to the sender and both networks are    
trained with the output of its partner. In case that they do not agree    
on the current output, $O^S O^R<0$, the weights of the    
sender/recipient are updated according to the following Hebbian learning      
rule \cite{5,11}.     
     
\begin{eqnarray}     
\mbox{if}  \  \Big( O^{S/R} y^{S/R}_k  > 0  \Big) \;     
& \mbox{then} &  \;     
w^{S/R}_{kj}   =  w^{S/R}_{kj}  - O^{S/R} \, x_{kj}  \nonumber \\     
\mbox{if} \ \Big( |w_{kj}^{S/R}| > L \Big)\; \; & \mbox{then} & \;    
 w_{kj}^{S/R} = \mbox{sign}(w_{kj}^{S/R}) \ L    
\label{two}    
\end{eqnarray}     
    
Only weights belonging to the one (or three) hidden units which are in    
the same state as that of their output unit are updated, in each one    
of the two networks.  Note that by using this dynamical rule, the    
sender is trying to imitate the response of the recipient and the    
recipient is trying to imitate the one of the sender.    
    
There are three main ingredients in our model which are essential    
for a secure key exchange protocol: First, from the knowledge of    
the output, the internal representation of the hidden units is not    
uniquely determined because there is a four fold degeneracy (for the    
output $+1$ there are four internal representations for the three    
hidden units $(1,1,1),~(1,-1,-1),~(-1,1,-1),~(-1,-1,1)$).  As a    
consequence, an observer cannot know which of the weight vectors is    
updated according to equation (\ref{two}). Second, we have chosen the    
parity machine since in the case of static weights, it is known that an    
opponent cannot obtain any knowledge about the rule if he is trained    
with less than $\alpha(L)N$ random examples (where $\alpha(L)=2.63$ for    
$L=3$ and for more details see \cite{12}).    
This analytic result    
favours the PM over other multilayer networks. Third, since each    
component is bounded by $L$, an observer cannot invert the sum of    
equation (\ref{two}); the network forgets \cite{13}. As a consequence    
of these three ingredients,    
the initial weight vectors cannot be recovered from the knowledge of    
the time dependent synchronized keys. All three of these mechanisms --    
hidden units, PM  as well as bounded weights -- make the problem hard for    
any observer.

\begin{figure}[ht]     
\centering      
\includegraphics[width=8cm]{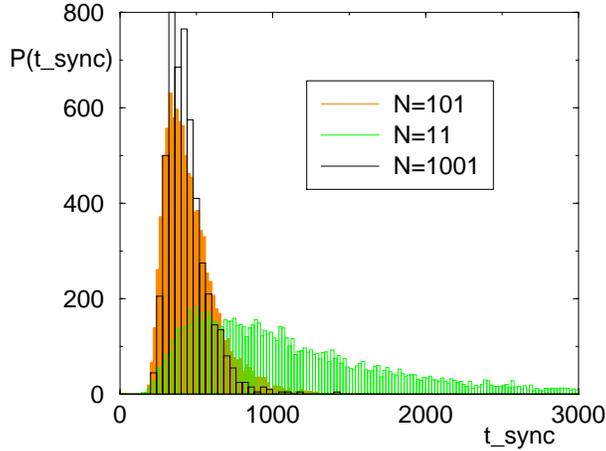}     
\caption{Distribution of synchronization time $t_{sync}$ for three sizes $N$      
of the two networks.     
         \label{Tsr}}     
\end{figure}

\begin{figure}[ht]     
\centering      
\includegraphics[width=8cm]{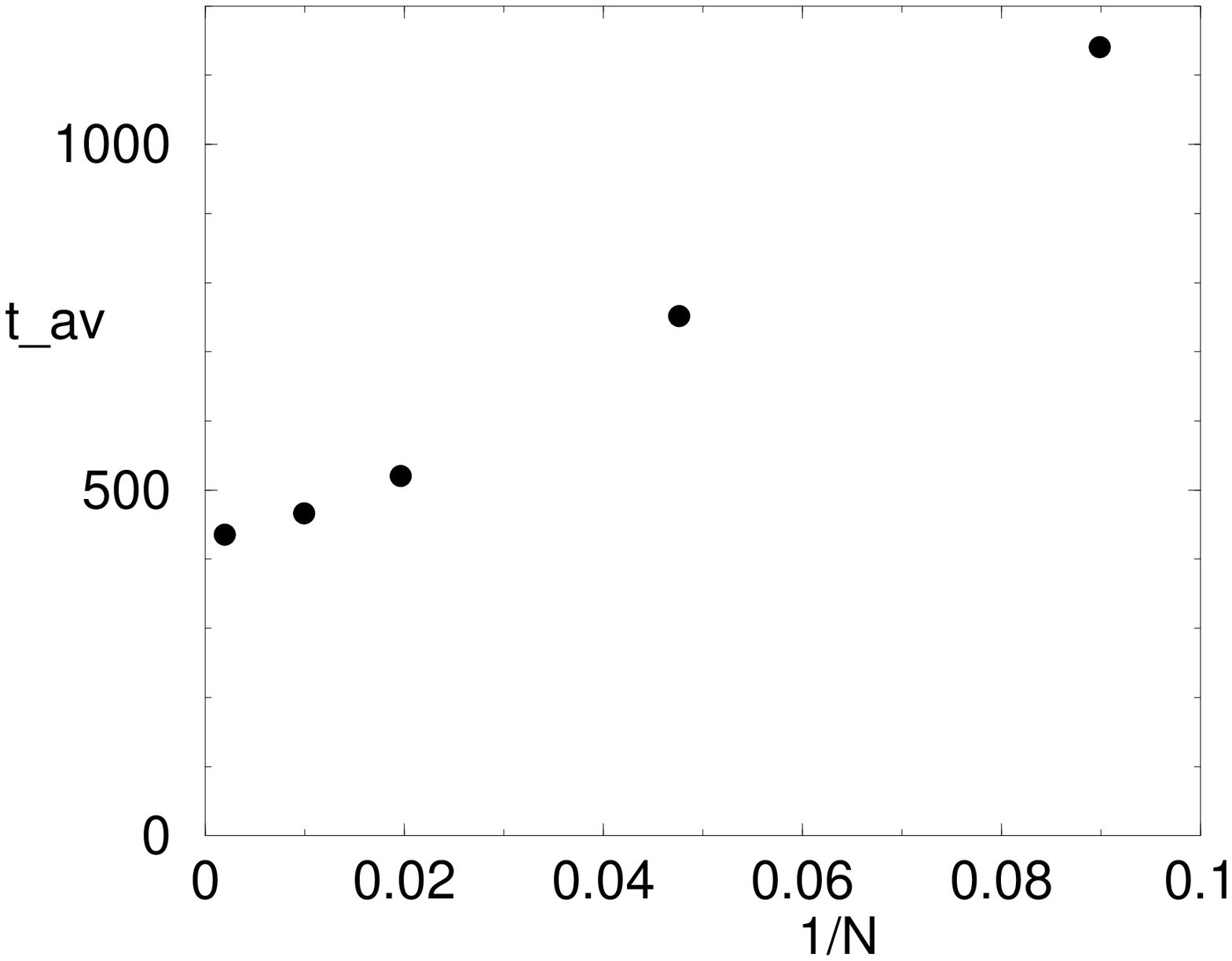}     
\caption{Average synchronization time as a function of $1/N$,    
for system size $N=11,21,51,101,1001$.     
         \label{tn}}     
\end{figure}

We find that the two PMs learning from each other are able to    
synchronize, at least for some parameters $K, L$ and $N$ \cite{14}.    
Our simulations show that after an initial relatively short    
transient time the two partners align themselves into antiparallel    
states.  It is easy to verify from our learning rule that as soon as    
the two networks are synchronized they stay so forever.  The number of    
time steps to reach this state depends on the initial weight vectors    
and on the sequence of random inputs, hence it is distributed. Fig.    
\ref{Tsr} shows the distribution of synchronization time obtained from    
at least $1000$ samples. It is    
evident that two communicating networks synchronize in a rather short    
time.  The average synchronization time $t_{av}$ decreases with    
increasing size $N$ of the system, see Fig. \ref{tn};     
it seems to converge     
to $t_{av} \simeq 410$ for infinitely large networks.     
Surprisingly,    
in the limit of large $N$ one needs to exchange only  about $400$ bits to    
obtain agreement between $3 N$ components. However, one should keep in mind     
that the two partners do not learn the initial weights of each other, they    
just are attracted to a dynamical state with opposite weight vectors.      
    
As soon as the weights of the sender and the recipient are    
antiparallel the public initialization of our private-key cryptosystem    
is terminated successfully and the encryption of the message starts.    
Now there are two possibilities to choose an algorithm: First,    
use a conventional encryption algorithm, for example a stream cipher    
like the well-known Blum-Blum-Shub bit generator \cite{1}. In this case    
the seed for this pseudo-random number generator is constructed from    
our weight vector after synchronization. 
Second, use the PM  
itself for a stream cipher  by multiplying its output bit with the     
corresponding bit of the secret data.

In the case of the PM, the complexity of the     
encryption/de\-cryp\-tion processes     
scales linearly with the size of the transmitted    
message, whereas the complexity of the     
synchronization process does not scale with the size of the network.    
Hence our construction is a linear    
cryptosystem \cite{15}.      
  
Now we examine a possible attack on our cryptosystem.  The opponent    
eavesdropping on the line knows the algorithm as well as the actual    
mutual outputs, hence he knows in which time steps the weights are    
changed.  In addition, the opponent knows the input $x_{kj}$ as well.    
However, the opponent does not know the initial conditions of the    
weights of the sender and the recipient. As a consequence, even for    
the synchronized state, the internal representations of the hidden    
units of the sender and the recipient are hidden from the opponent and    
he does not know which are the weights participating in the learning    
step.  For random inputs all four internal representations    
appear with equal probability in any stage of the dynamical process,    
hence for $t$ training steps there are $4^t$ possibilities to select    
internal representations.

Therefore, on the time scale of synchronization the observing network    
has no chance to obtain complete knowledge about the other two    
networks.  We have simulated  attacks of an observer, assuming that    
the most effective algorithm is a network which     
has identical architecture to the    
recipient: A PM with the same  learning    
rule and parameters as described above. The observing network is    
trained with the input vector and output bit of the sender, and the    
training step (\ref{two}) is performed only if sender and recipient    
disagree with each other. 
Note that the sender does not react to the
output of the opponent, which results in a large noise to signal ratio
compared with the recipient \cite{ref2}.     
    
\begin{figure}[ht]     
\centering      
\includegraphics[width=8cm]{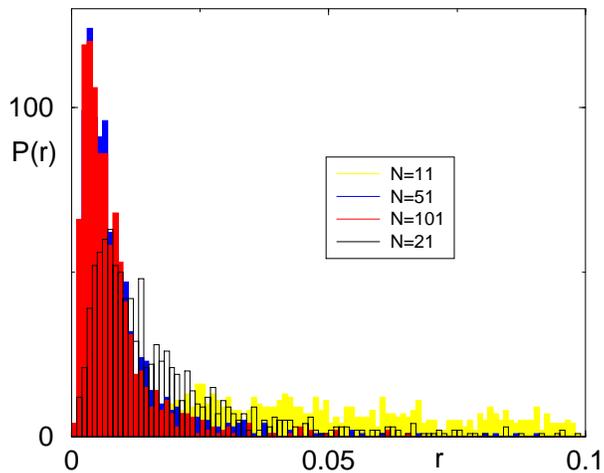}     
\caption{Frequency of the ratio $r$ between    
synchronization and learning times.    
         \label{rat}}     
\end{figure}

The learning rule (\ref{two}) may be considered for each component of    
the weight vectors as a kind of biased random walk with reflecting    
boundaries.  Therefore, for very long times, the observer may take the    
weight vector of the other network by chance. The distribution of the    
ratio between the time $t_{sync}$ the sender and recipient need to    
synchronize and the learning time $t_{learn}$ the opponent needs for    
complete overlap is shown in Fig. \ref{rat}. For $N=101$ the average    
learning time is a factor of about $125$ larger than the corresponding    
synchronization time. In addition, with increasing system size the tail    
of the distribution for larger ratios is reduced.

Hence the time to synchronize by chance is very long and in the    
example discussed here it is of order $O(10^5)$ \cite{17}.      
The heart of our    
cryptosystem is that synchronization is a much simpler task than    
tracking by an observer.    
    
This principle is also supported by the following observation. Assume    
that the observer has already some knowledge about one of the    
networks, he knows $M$ out of $N$ components for each of the    
  three hidden units.  We have measured whether the observer succeeds    
to synchronize within 2000 time steps -- a time much longer than the    
average synchronization time for the two communicating networks.  For    
$M=N$ the observer is parallel to one of the networks and remains so    
forever.  But already for $M=N-1$ we find a high probability that the    
weight vector of the observer separates completely from its almost    
parallel alignment. For smaller values of $M$ this probability    
decreases fast to zero. Surprisingly, even in case the observer has    
almost complete knowledge about the two partners, he does not succeed    
to achieve complete information from learning examples.  This fact    
reduces the probability of the opponent to imitate one of the    
communicating networks using an ensemble of PMs.

Our key exchange protocol can be generalized to include 
{\it Bit-Packages} as is briefly described below.    
An important issue for the implementation of our    
cryptosystem is to accelerate the synchronization process from    
hundreds of time steps to  a few dozens while    
keeping the security of our channel. Surprisingly, both of these two    
goals can be achieved simultaneously sending bit packages (BP). In    
this scenario the process contains the following steps: (a) The sender    
and the recipient generate $B>1$ common inputs.  (b) The sender and    
the recipient calculate the output of their PM for the set of $B$    
inputs and store the $B$ sets of the corresponding values $y_{ki}\;    
(i=1, ...,B)$ of the hidden units (the internal representations) (c) The    
transmission of mutual information; the sender/recipient sends a    
package consisting of $B$ bits ($b_i^{S/R}$) to the recipient/sender.    
(d) The sender and the recipient are updating their weights using the    
same learning rule as for $B=1$:  In case that bit $b^S_i \ne b^R_i$    
the learning process is taking place as before using the corresponding    
internal representations. The synchronization time is dramatically    
reduced, as is shown in Fig. 5.  For instance, for $N=21$, $K=3$,    
$L=3$, synchronization is achieved after 12 bit packages if the size    
of the package is larger than $B \ge 32$.    
    
\begin{figure}     
\centerline{\epsfxsize=3.25in \epsffile{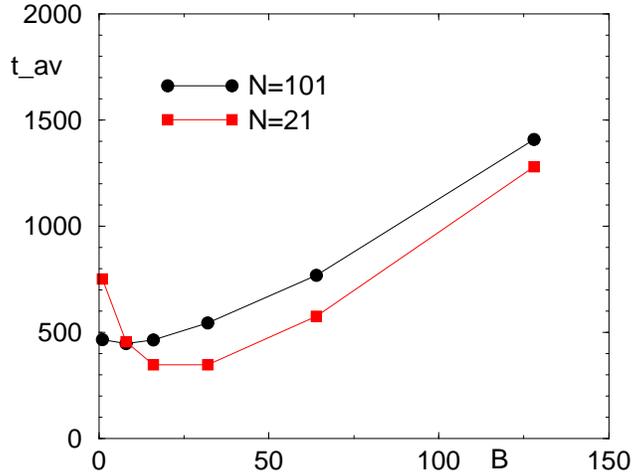}}     
\caption{Total number of transmitted bits until synchronization. $B$      
  is the number of bits in each bit package exchanged between    
    sender and recipient.    
         \label{tbp}}     
\end{figure}    

Other extensions of our method as well as the analytical calculation of
the distribution of $t_{sync}$, $t_{learn}$ for various $K, L$ and
continuous weights, and version space of the PMs which are
consistent with the training set will be dicussed in \cite{ref2}.

Finally, we want to remark that synchronization is a subject of recent    
research in neuroscience, where for instance, in experiments on cats    
and monkeys one has found that the spike activity of neurons in the    
visual cortex has correlations which depend on the kind of optical    
stimulus shown to the animal \cite{18}. The phenomenon described    
here suggests that synchronization can be used by biological neuronal    
networks or by networks of the immune system to exchange secure    
information between different parts of an organism.


\begin{thebibliography}{10}
\bibitem{15} As for the hardware implementation of our    
cryptosystem, the calculation of the    
local fields, eq. (1), are similar to a linear filter which can be    
implemented in parallel using existing technologies.  Hence, the    
complexities of the synchronization process is of $O(1)$ only.
\bibitem{16} Another possible solution is based on continuous weights.
A similar synchronization was observed for a PM with continuous
weights, where the norm of the weights is set equal to $1$ after each
learning step \cite{14,ref2}.
\bibitem{17} Note that the average learning time for $N=101$ scales
exponentially with $L$ and for $L=1,2,3,4,$ was found in simulations
to be equal to $58,1480,90380,\sim 10^7$.
\end{thebibliography}
\end{document}